\documentclass[reprint,amsmath,amssymb,aps,prl,showkeys,superscriptaddress]{revtex4-2}

\usepackage{graphicx}
\usepackage{bm,color}
\usepackage{booktabs}
\usepackage{array}
\usepackage{braket}
\usepackage{dsfont}
\usepackage[T1]{fontenc}
\allowdisplaybreaks

\newcommand{\Tr}{\operatorname{Tr}}

\newcommand{\ew}[1]{\langle {#1} \rangle}
\newcommand{\kb}[2]{| {#1} \rangle \langle {#2}|}
\newcommand{\rjm}{\rho_{j,m}}

\begin{document}

\title{Signatures of superradiance as a witness to multipartite entanglement}%

\author{Frederik Lohof}
\affiliation{Institute for Theoretical Physics and Bremen Center for Computational Material Science, University of Bremen, 28359 Bremen, Germany}

\author{Daniel Schumayer}
\affiliation{Dodd-Walls Centre, Department of Physics, University of Otago, Dunedin 9016, New Zealand}

\author{David A. W. Hutchinson}
\affiliation{Dodd-Walls Centre, Department of Physics, University of Otago, Dunedin 9016, New Zealand}
\affiliation{Centre for Quantum Technologies, National University of Singapore, Singapore 117543}

\author{Christopher Gies}
\affiliation{Institute for Theoretical Physics and Bremen Center for Computational Material Science, University of Bremen, 28359 Bremen, Germany}

\date{\today}

\begin{abstract}
    Generation and detection of entanglement is at the forefront of most quantum information technologies.
    There is a plethora of techniques that reveal entanglement on the basis of only partial information about the underlying quantum state, including entanglement witnesses.
    Superradiance refers to the phenomenon of highly synchronized photon emission from an ensemble of quantum emitters that is caused by correlations among the individual particles and has been connected by Dicke himself to the presence of multipartite entangled states.
    We investigate this connection in a quantitative way and discuss, whether or not signatures of superradiance from semiconductor nanolasers, manifesting themselves as a modification of the spontaneous-emission time, can be interpreted as a witness to detect entanglement in the underlying state of the emitters.
\end{abstract}


\maketitle

Radiative coupling between localized emitters has been of recurrent interest in different research areas over the past decades \cite{dicke_coherence_1954, eberly_superradiance_1972,gross_superradiance_1982,scheibner_superradiance_2007,scully_super_2009,hotter_cavity_2023}.
In a simple picture, coupling to a common light field can cause spatially distant emitters to align dipoles, resulting in correlated emission phenomena \cite{leymann_sub-_2015}.
Well known effects that arise from this are sub- and superradiance -- correlation-induced modifications of spontaneous emission due to the presence of other emitters that can enhance or reduce the spontaneous emission rate, respectively.
While coupling via the free-space radiation field requires emitters to be close together \cite{dicke_coherence_1954} or regularly positioned \cite{masson_universality_2022} for the effect to occur, these conditions can be relaxed if emitters couple to a common mode of a microcavity.
This situation has been investigated in the past, and signatures of sub- and superradiant emission have been demonstrated in ensembles of quantum dots \cite{scheibner_superradiance_2007,jahnke_giant_2016, kreinberg_emission_2017}, quantum wells \cite{timothy_noe_ii_giant_2012}, and superconducting circuits \cite{mlynek_observation_2014,wang_controllable_2020}.
Dicke described the phenomenon of sub- and superradiance in terms of collective states that are eigenstates equivalent to those of the total angular momentum.
The maximal angular momentum states of $N$ emitters with $j = N/2$ are known today as Dicke states and exhibit a quadratic dependence of their emission rate with the number of particles.
They are multipartite entangled quantum states that can act as a resource for a range of different quantum information applications \cite{prevedel_experimental_2009,pezze_quantum_2018,miguel-ramiro_delocalized_2020}.
Dicke's picture allows a somewhat different access to explaining the modification of spontaneous emission time in terms of dark and bright states \cite{dicke_coherence_1954,temnov_superradiance_2005,auffeves_few_2011}.
Many consider the term \emph{superradiance} to be bound to the emission from a Dicke state, whereas spontaneously created dipole correlations that also modify spontaneous emission in a similar manner are referred to as \emph{superfluorescence} \cite{bonifacio_cooperative_1975,vrehen_superfluorescence_1980,cong_superfluorescence_2015}.
In a reciprocal way, it is not clear in how far altered spontaneous emission behavior is indicative for the presence of Dicke states \cite{toth_detection_2007}.

\begin{figure}[t]
    \centering
    \includegraphics[width=1.\linewidth]{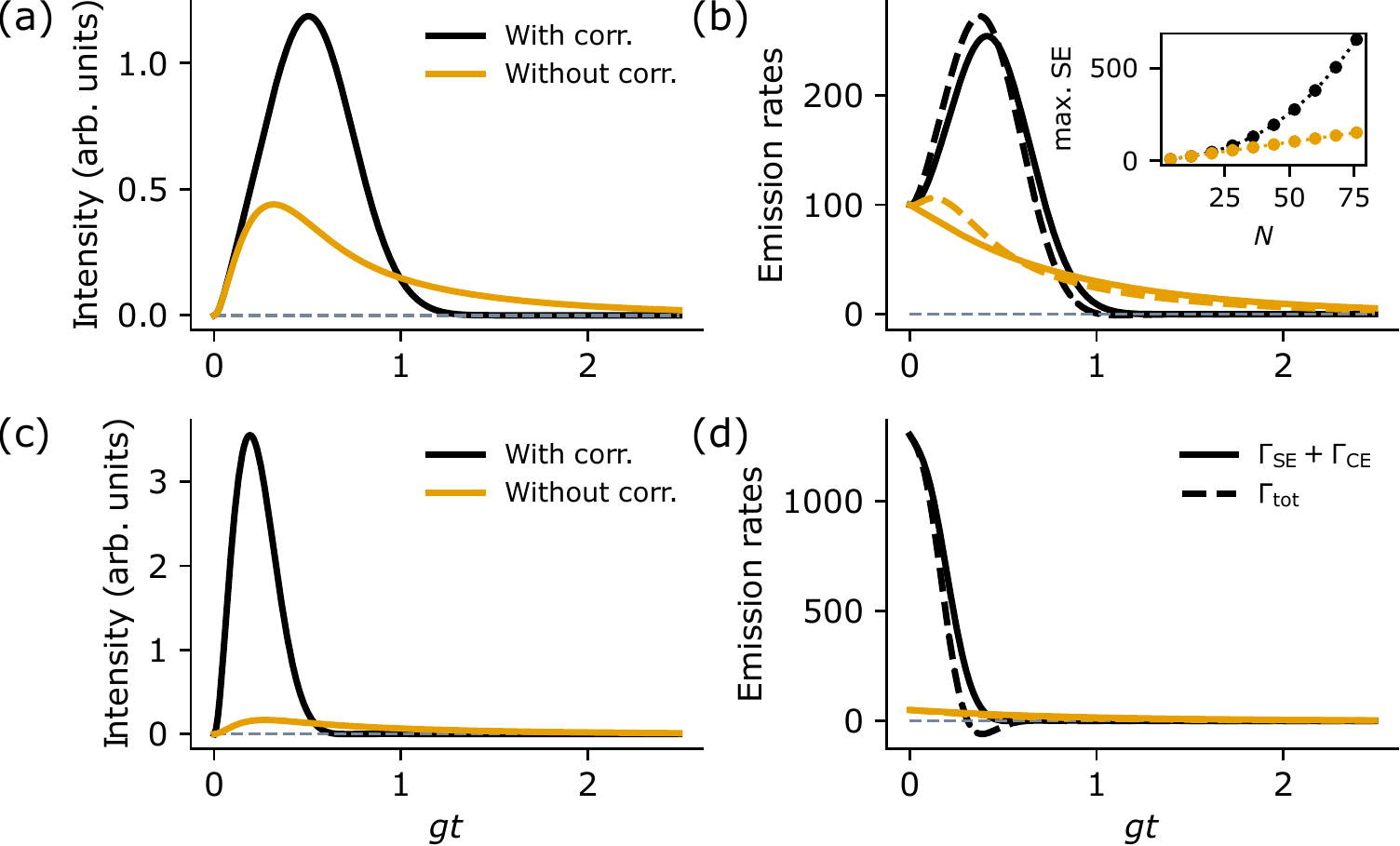}
    \caption{(a) Comparison of emission from initially fully inverted emitters ($N=50$) with (black) and without (orange) emitter correlations.
        (b) Combined spontaneous emission $\Gamma_\mathrm{SE}$ with superradiant enhancement $\Gamma_\mathrm{SR}$ (solid lines).
        Without emitter correlations $\Gamma_\mathrm{SR}=0$.
        Total emission $\Gamma_\mathrm{tot}$ (dashed) includes stimulated emission as well.
        Inset: Emitter-number dependence of the maximum   spontaneous emission rate.
        (c)-(d) Same as above, but with the emitter ensemble initially in the half inverted Dicke state $\ket{D_{N,N/2}}$.
        Parameters are $\gamma/g=1$, $\kappa/g = 20$.}
    \label{fig:SR_emission_enhancement}
\end{figure}
The purpose of this letter is to establish and quantify the connection between correlation-induced spontaneous emission-time modification and multi-partite entanglement in terms of Dicke states.
We consider quantum-dot based cavity-QED systems, but the interpretation and consequences apply to other systems as well.
Determining entanglement in a multi-partite system is not trivial and no general measures apply \cite{guhne_entanglement_2009}.
Here, we connect superfluorescent intensity bursts to entanglement witnesses for Dicke states that are related to electronic structure factors \cite{krammer_multipartite_2009}.
This is a particularly suitable approach, as this formulation of the entanglement witness directly relates to electronic correlation functions that are responsible for changes of the spontaneous-emission time, and which are accessible by equation-of-motion techniques used in semiconductor quantum optics \cite{kira_semiconductor_2011,gies_semiconductor_2007,leymann_sub-_2015}.
As we show, this facilitates a natural connection between semiconductor quantum-optical methods and methods from quantum-information theory.

In the past, the presence of superradiant spontaneous emission enhancement has been seen as an indicator for Dicke superradiance.
Here, we present proof that in the superradiant regime of nanolasers, the emitter ensemble rarely transitions through an intermediate Dicke state with a high degree of entanglement, but instead a coexistence of sub- and superradiant states is observed during light emission inhibiting the occurrence of detectable entanglement.
We highlight how the formation of emitter correlations and the $N^2$ scaling of the spontaneous emission rate is a necessary but not sufficient condition for the presence of multipartite entanglement during the emission process.

\paragraph*{Superradiant emission and entanglement.}---
Superradiance is caused by correlations between emitters and can arise in an ensemble of $N$ emitters coupling to the common electromagnetic field.
While superradiance has been associated with highly directional emission, e.g.~in the case of atomic vapors emitting into free space \cite{eberly_superradiance_1972}, the situation is different if the emitters are confined to a microcavity and couple to a common mode.
In this case, the spatial profile of the emission is solely determined by the cavity mode, and interference effects, such as due to inhomogeneity, play no role.

We use a model of $N$ individual two-level emitters coupled to a single bosonic mode governed by the Tavis-Cummings Hamiltonian
\begin{equation}
    H= \sum_i \frac{\omega_q }{2} \, \sigma_i^z+ g \sum_i (\sigma_i^+ a + \sigma_i^- a^\dag )  + \omega_c\, a^\dag a \, ,\label{eq:H_RWA}
\end{equation}
where $\omega_q$ and $\omega_c$ are the emitter and cavity resonance frequencies ($\hbar=1$), and $g$ is the light-matter interaction strength.
The parafermionic operators $\sigma_i^z$ and $\sigma_i^\pm = \frac{1}{2}(\sigma_i^x \pm \mathrm{i}\sigma_i^y)$ are the inversion, raising and lowering operators for emitter $i$ \cite{wu_qubits_2002}, while $a^\dag$ and $a$ are bosonic operators creating and annihilating photons in the cavity mode.
We consider an open quantum system \cite{breuer_theory_2007} subject to emitter decay, photon loss, and pure dephasing at rates $\gamma$, $\kappa$, and $\gamma_\phi$.
Throughout the paper we use a pure dephasing of 20$\,\mu\mathrm{eV}$ typical for low-temperature III/V semiconductor quantum dots.
We apply the cluster-expansion technique up to doublet level \cite{gies_semiconductor_2007,kira_semiconductor_2011} taking into account pair correlations between individual emitters.
This well-established approach provides access to the dynamics of fundamental quantities, such as the output intensity, for large emitter ensembles in the weak coupling regime ($g^2/(\gamma + \kappa) \ll 1$).
All used equations are derived in the Supplementary Information (SI).
The temporal evolution of the photon number in the cavity $n(t)$ can be written as
\begin{equation}
    \dot{n}(t) = \Gamma_\mathrm{SE}(t) + \Gamma_\mathrm{StE}(t) + \Gamma_\mathrm{CE}(t) - \kappa n(t),
\end{equation}
where the individual terms are the spontaneous emission (SE) rate $\Gamma_\mathrm{SE} = I_0 \frac{N}{2}(1+\ew{\sigma_z})$, stimulated emission (StE) $\Gamma_\mathrm{StE} = I_0 Nn\ew{\sigma_z}$, and the correlated emission (CE) responsible for sub- and superradiant SE-time modifications $\Gamma_\mathrm{CE} = I_0 N(N-1)C_0$.
Here, $I_0=\frac{4g^2}{\kappa + \gamma + 2 \gamma_\phi}$ is the single-emitter emission rate into the cavity, $\ew{\sigma_z}$ is the inversion, and $C_0 = \ew{\sigma_i^+ \sigma_j^-}$ are emitter pair correlations.
Importantly, the cluster-expansion approach allows us to switch off and on these correlation contributions to asses their influence.
\begin{figure}[t]
    \centering
    \includegraphics[width=1.\linewidth]{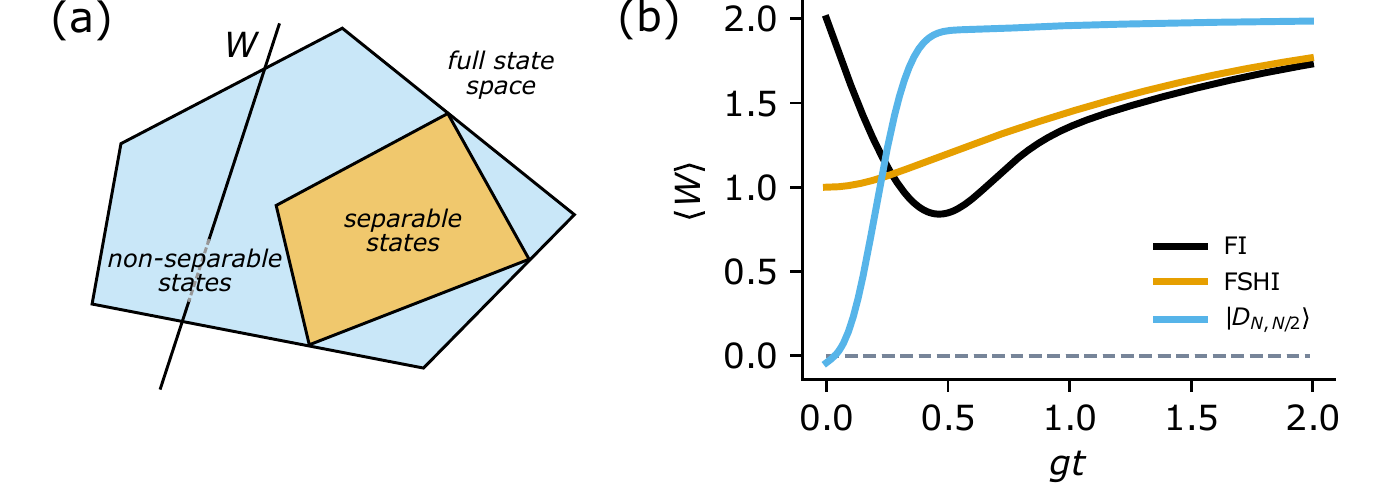}
    \caption{(a) Schematic representation of the space of all physical states $\rho$ with all separable states as a convex subset (orange).
The witness $W$ defines a hyperplane given by $\Tr[W \rho]=0$, bisecting the whole space into detected and non-detected states.
        (b) Structure factor witness $\ew{W}$ evaluated for different initial conditions: fully inverted (FI) emitters (black), fully-separable half-inverted (FSHI) product state (orange), and half-inverted Dicke state $\ket{D_{N,N/2}}$ (blue).
Parameters are identical to Fig~\ref{fig:SR_emission_enhancement}.}
    \label{fig:SR_witness_EoM}
\end{figure}

Fig.~\ref{fig:SR_emission_enhancement}(a) shows the output intensity of an initially fully inverted system of $N=50$ emitters with (black) and without (orange) the influence of emitter correlations as a function of time.
The well-known fingerprint associated with superfluorescence is apparent as an emission burst, during which most photons are emitted within a shortened time interval.
Fig.~\ref{fig:SR_emission_enhancement}(b) shows the corresponding emission rates.
Without emitter correlations, the SE rate decays exponentially, while the presence of emitter correlations  strongly enhances the emission rate before it falls beneath the uncorrelated value -- these regimes are referred to as super- and subradiant.
To verify that we truly are in a regime dominated by spontaneous emission, dashed lines show the total emission including stimulated contributions, which are indeed small.
In Fig.~\ref{fig:SR_emission_enhancement}(c)-(d) we also show the emission output and emission rates from an ensemble initialized in the half-inverted Dicke state $\ket{D_{N, N/2}}$ at $t=0$.
Emission from this highly entangled state is eponymous for Dicke superradiance: strongly correlated and known to possess the largest SE-rate enhancement.
The emission rate for Dicke states $\ket{D_{N,k}}$ can be given in analytic form, $\Gamma_\mathrm{CE} = I_0(N-k)k = I_0N^2(1-\ew{\sigma_z}^2)/4$, showing the proportionality to $N^2$ at a fixed inversion $\ew{\sigma_z}$.
Without correlations, the SE rate is only linear in $N$: $\Gamma_\mathrm{SE} = I_0 k = I_0 \frac{N}{2}(1 + \ew{\sigma_z})$  (see the SI).

We now turn to the central question: To what extent is superradiant SE enhancement indicative for entanglement between the correlated emitters?
While for the half-inverted Dicke state in Fig.~\ref{fig:SR_emission_enhancement}(c),(d), entanglement is present by design, the situation is less clear for the temporal evolution of an initially fully inverted system.
Our results in Fig.~\ref{fig:SR_emission_enhancement}(a),(b) show that at half inversion, the SE rate is strongly enhanced, and the inset to panel (b) reveals that this increase indeed scales as $N^2$, suggesting the transitory presence of the entangled half-inverted Dicke state.
To answer this question, we employ an entanglement witness that is related to structure factors \cite{krammer_multipartite_2009} 
\begin{equation}
    W = \mathds{1} - \binom{N}{2}^{-1} \sum_{i<j} \left( c_x \sigma_i^x \sigma_j^x + c_y \sigma_i^y \sigma_j^y+ c_z \sigma_i^z \sigma_j^z\right)\label{eq:SF_witness}
\end{equation}
with real coefficients $|c_{x/y/z}| \leq 1$ and the binominal coefficient $\binom{n}{k}$.
As illustrated in the schematic picture of Fig.~\ref{fig:SR_witness_EoM}(a), the witness defines a hyperplane given by $\Tr[W \rho]=0$, dividing the space of all physical states into two parts, one ($\Tr[W \rho]\geq 0$) containing the convex subset of all separable states (orange), and one ($\Tr[W \rho]<0$) containing all states that are said to be detected by the witness.
Thus, a positive value for a witness is a necessary, but not a sufficient criterion for a state to be separable \cite{guhne_entanglement_2009}.
It has been shown that for $c_x=c_y = 1$ and $c_z = -1$, $W$ detects states close to the half-inverted Dicke state $\ket{D_{N, N/2}}$ \cite{krammer_multipartite_2009}.
We note that there are other witnesses that are more sensitive in detecting entangled states close to Dicke states \cite{guhne_entanglement_2009}.
However, these typically contain higher-order correlations that are not accessible by the equation-of-motion approach, which enables us to evaluate $\langle W\rangle$ even for large emitter ensembles.
Due to the symmetry of the Hamiltonian under exchange of emitters, the expectation values $\ew{\sigma_i^\alpha}$ and $\ew{\sigma_i^\alpha \sigma_j^\alpha}$ do not depend on the particle indices $i,j$.
In this case, we obtain from Eq.~\eqref{eq:SF_witness} 
\begin{equation}
    \ew{W} = 1 - 4\mathrm{Re}[C_0] + C_{zz}
\end{equation}
with  $C_{0} = \ew{\sigma_i^+ \sigma_j^-}$ and $C_{zz} = \ew{\sigma_i^z \sigma_j^z}$ ($i\neq j$).
It is a key insight of this letter that this particular witness is directly connected to the correlated-emission term $\Gamma_{\mathrm{CE}}$, which contains the same type of emitter-pair correlations and is, therefore, directly accessible by the cluster-expansion approach.
In fact, this connection has been one of the motivations for investigating the connection between entanglement and superradiant spontaneous-emission rate enhancement.

Fig.~\ref{fig:SR_witness_EoM}(b) shows $\ew{W}$ for different initial conditions of the previous ensemble of 50 emitters.
For initially fully inverted emitters (black), there is a significant dip in the witness expectation value during the emission that coincides with the maximal intensity output in Fig.~\ref{fig:SR_emission_enhancement}(a).
In the past, the initial burst of emission after excitation of the system was connected to the presence of a Dicke state at the half-inversion point of the emission.
Here, the fact that the witness stays positive shows that this is \textit{not} the case.
As a comparison, $\ew{W}$ is evaluated for a fully-separable half-inverted product state (orange) $\rho_{\frac{1}{2}}^{\otimes N}$ with $\rho_{\frac{1}{2}} = \frac{1}{2}(\kb{0}{0}+\kb{1}{1})$, and for the half-inverted Dicke state (blue) $\rho_{N,N/2} = \kb{D_{N,N/2}}{D_{N,N/2}}$.
Both states have the same initial excitation level $\ew{\sigma_z}=0$.
While no entanglement is witnessed for the separable initial state, the Dicke state is detected by construction of the witness and starts out with the minimal possible value of $\ew{W} = -2/(N-1)$ \cite{krammer_multipartite_2009}.

The fact that Dicke states close to half-inversion are not detected during the emission despite the observation of a superradiant spontaneous-emission rate enhancement $\propto N^2$ is a key result of this letter.
It originates from the fact that entanglement, as we infer from the definition of $\langle W \rangle$, mainly arises from the real part of the emitter-emitter correlations $C_0$, whereas the SE rate modification $\Gamma_\mathrm{CE}$ contains an additional factor of $N(N-1)$, reflecting that SE is enhanced by all pairwise correlations within the ensemble. Entanglement does not benefit from this combinatorial factor.

 
%
\begin{figure}[t]
    \centering
    \includegraphics[width=1.\linewidth,trim={0.cm 0.cm .0cm 0.0cm},clip]{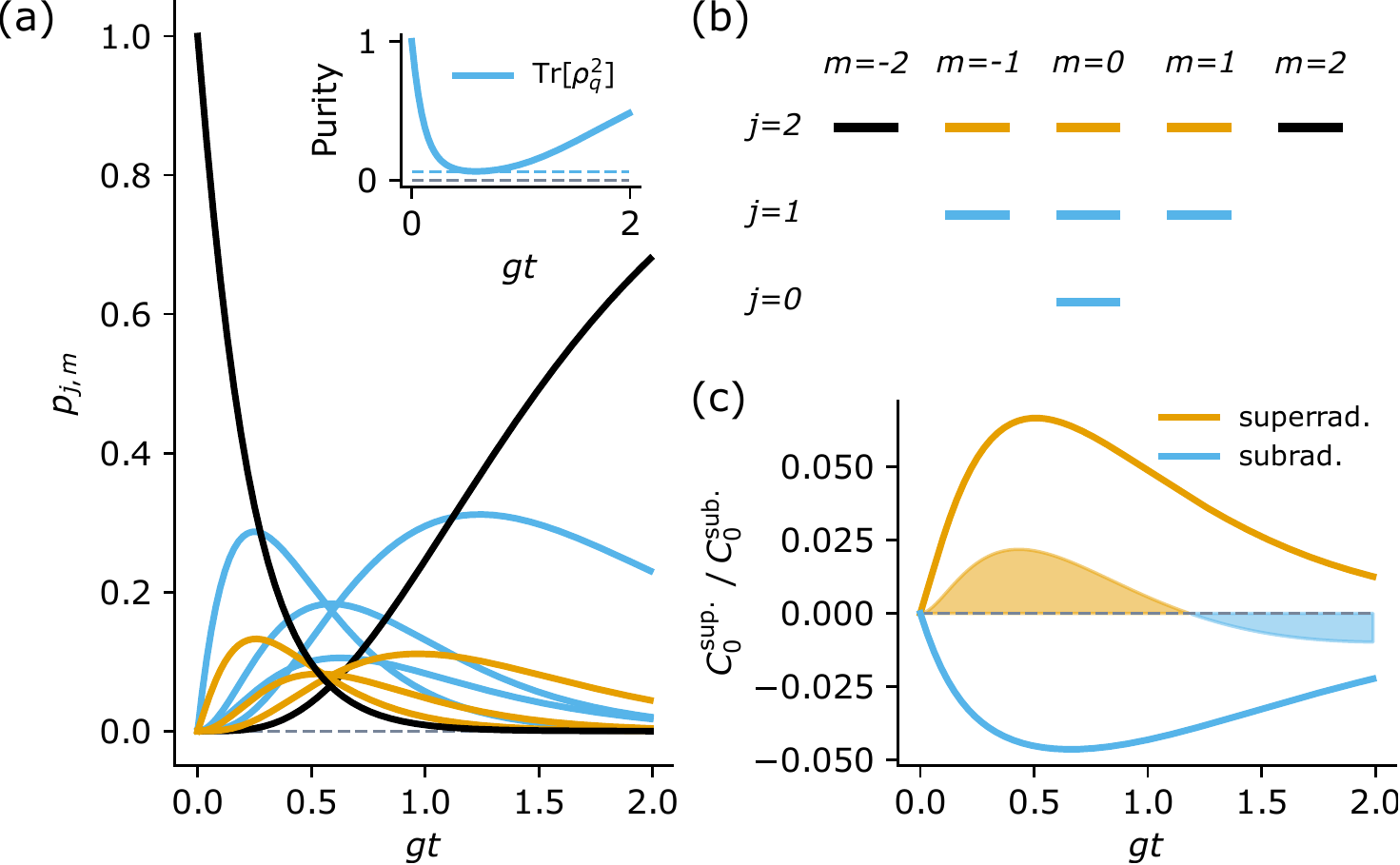}
    \caption{
        (a) Decomposition of the reduced 4-emitter state into components $\rjm$ according to Eq.~\eqref{eq:rho_decomposition}.
        Colors indicate competing superradiant (orange), subradiant (blue) contributions as well as the separable ground and inverted state (black).
        The inset shows the state purity $\Tr[\rho_q^2]$ showing the transitive mixing of angular momentum components.
        In (b) the corresponding structure of possible angular momentum numbers ($j,m$) is shown, where the coloring reflects the contributions shown in (a).
        (c) Emitter correlation $C_0$ separated into sub- and superradiant components as given in Eq.~\eqref{eq:C_0_sup}.
        The shaded area indicates the sum of the two competing effects. Parameters are $N=4$, $\kappa/g = 20$, $\gamma/g = 1$.
    }
    \label{fig:perm_inv_decomposition_and_negativity}
\end{figure}
\paragraph*{Pair correlations do not equal entanglement.}
---If no Dicke state is detected during the emission, the question remains what correlated state is actually formed instead.
To investigate this, we analyze a system of fewer emitters that is amenable to a treatment by master equations for the system's density operator.
We numerically integrate the  Lindblad-von Neumann  equation for an ensemble of four emitters.
Given that we start with a permutationally invariant initial state, it can be shown \cite{xu_simulating_2013, shammah_open_2018} that the dynamics only couples permutationally invariant emitter states that are characterized by fixed (pseudo-) spin angular momentum eigenstates of the collective spin operators $\bm{J}^2$ and $J_z$, where
\begin{equation}
    J_z = \frac{1}{2}\sum_i \sigma_i^z.
\end{equation}
As a consequence we can decompose the marginal state of the emitters as
\begin{equation}
    \rho_\mathrm{q}(t) = \sum_j \sum_{m=-j}^j p_{j,m}(t) \rjm,\label{eq:rho_decomposition}
\end{equation}
where $(j,m)$ label the permutationally invariant components
\begin{equation}
    \rjm = \frac{1}{d_{jm}}\sum_{i=1}^{d_{jm}} \kb{j,m,i}{j,m,i},\label{eq:rho_jm}
\end{equation}
and $\ket{j,m,i}$ the eigenstates with $\bm{J}^2\ket{j,m,i} = j(j+1)\ket{j,m,i}$ and $J_z\ket{j,m,i} = m\ket{j,m,i}$.
The factor $d_{jm} = \frac{2j+1}{N+1}\binom{N+1}{\frac{N}{2}-j}$ is the degree degeneracy of the common eigenspace of $\bm{J}^2$ and $J_z$ and is equal to the multiplicity of the irreducible representation $(j,m)$ in the Clebsch-Gordan decomposition series for adding $N$ (pseudo-) spin-$\frac{1}{2}$ systems.
The quantum number $i$ is introduced to lift this degeneracy.
Note that $j_\mathrm{min} = 0$ ($\frac{1}{2}$) for even (odd) number of emitters, $j\leq \frac{N}{2}$ and $-j \leq m \leq j$.

In the following, we use the orthogonality relation 
\begin{equation}
    \Tr [\rjm \rho_{j',m'}]= \frac{1}{d_{jm}} \delta_{jj'} \delta_{mm'}
\end{equation}
to extract the coefficients $p_{j,m}$ from any given system state $\rho$.
In this notation, Dicke states are those with maximal $j=\frac{N}{2}$ with $|m|<j$ and have multiplicity $d_{jm}=1$.
Fig.~\ref{fig:perm_inv_decomposition_and_negativity}(b) gives a representation of all possible states, of which the Dicke states are marked in orange.
Fig.~\ref{fig:perm_inv_decomposition_and_negativity}(a) shows the time dependence of the coefficients $p_{j,m}$ during the superradiant emission pulse.
We see subsequent occupation of the entangled Dicke states during the emission (orange), as well as large contributions from the separable, fully inverted state $\ket{j,j}$ and ground state $\ket{j,-j}$ (black).
Furthermore, dissipative processes in the Lindblad-von Neumann master equation create contributions with lower total angular momentum $j<N/2$ (blue), which are subradiant.

Here, the main distinction between SE-rate modification and creation of entanglement is revealed.
All permutationally invariant components $\rjm$ of the emitter state (except for fully inverted and ground state) are non-separable, as can be shown by various entanglement measures such as the negativity (see the SI).
However, the strong mixing of the entangled components leads to an overall separable state (i.e.~entanglement is \textit{sublinear} \cite{pittenger_convexity_2002, guhne_entanglement_2009}).
The mixing is also reflected in the low purity $\Tr[\rho_q^2]$ during the emission pulse shown in the inset of Fig.~\ref{fig:perm_inv_decomposition_and_negativity}(a).
In contrast, all SE-rate modifications considered here result from \textit{linear} operator expectation values, allowing to analyze the emission rates in terms of linear combinations of the individual components with fixed $(j,m)$.
The emitter correlations for any $\rjm$ can be calculated explicitly and are given by (see the SI)
\begin{equation}
    C_0(j,m)=\frac{j(j+1)-m^2 -\frac{N}{2}}{N(N-1)} \label{eq:C0_jm}
\end{equation}
and
\begin{equation}
    C_{zz}(j,m)=\frac{4 m^2 - N}{N(N-1)}.
\label{eq:Czz_jm}
\end{equation}
By separating all components $\rjm$ according to their values of $C_0(j,m)$, the time evolution of the purely superradiant component can be calculated and is given by 
\begin{equation}
    C_0^\mathrm{sup.} (t) = \sum_{\substack{j,m\\ C_0(j,m)>0}} p_{j,m}(t) C_0(j,m), \label{eq:C_0_sup}
\end{equation}
with the subradiant component $C_0^\mathrm{sub.} (t)$ analogously.
The colors in all panels of Fig.~\ref{fig:perm_inv_decomposition_and_negativity} are chosen to reflect the sub- (blue) or superradiant (orange) nature of the individual contributions.
In Fig.~\ref{fig:perm_inv_decomposition_and_negativity}(c) we show how the sub- and superradiant contributions change during the emission pulse.
The result of this competition between sub- and superradiant contributions is represented by the shaded curve, which is the sum of the two contributions.
It shows how the observed superradiant emission burst in Fig.~\ref{fig:SR_emission_enhancement}(a,b) arises from competing effects, with net superradiant behavior in the beginning that later transitions to a net subradiant behavior.

Finally, we address the stability of our results in the presence of inhomogeneous broadening of the emitter transition frequencies that is present in ensembles of epitaxially grown quantum dots.
It has previously been argued that the cavity acts as a spectral filter, decoupling all non-resonant emitters from the resonant part of the ensemble [ref].
In this sense, the emitter number $N$ in our model is an effective quantity describing only those emitters that are close to resonance with the cavity mode.
The SI provides a deeper discussion and a calculation for inhomogeneous emitters.

\begin{figure}[t]
    \centering
    \includegraphics[width=1.\linewidth,trim={0.cm 0.cm .0cm 0.0cm},clip]{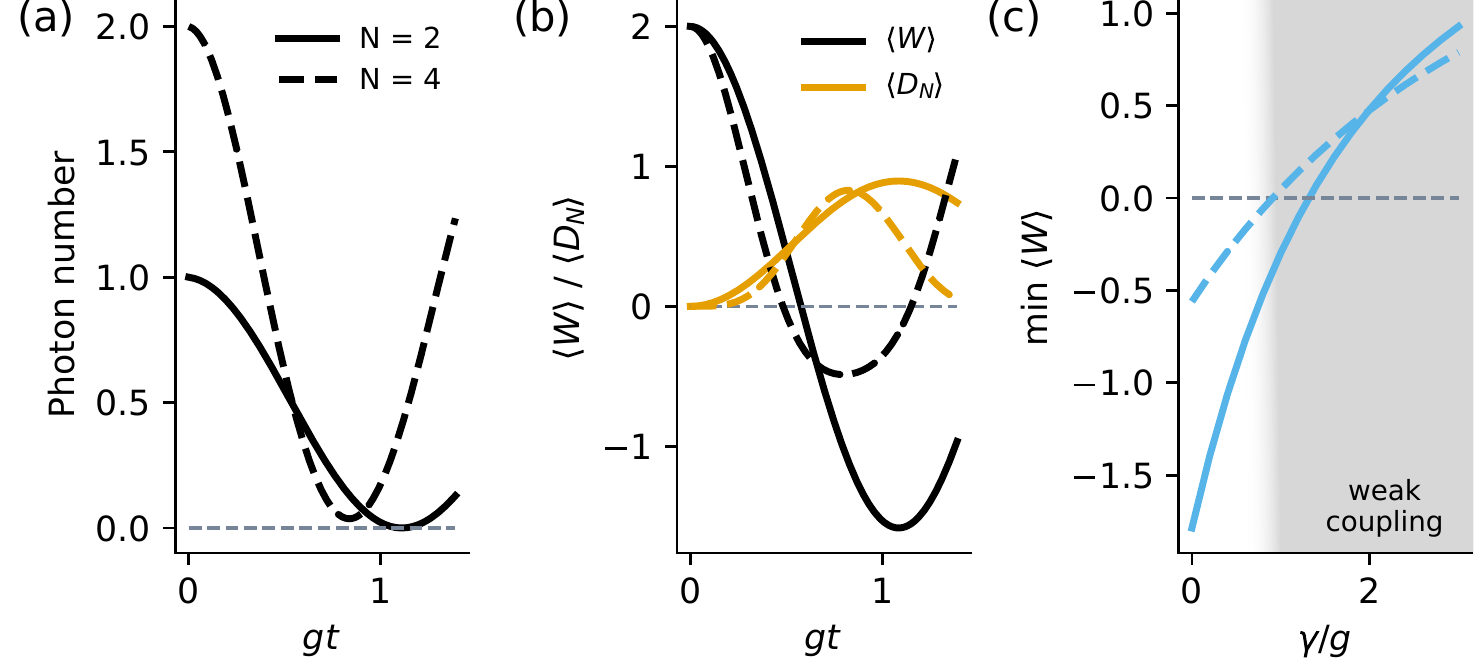}
    \caption{
        Realization of Dicke states in the strong coupling regime for $N = 2$ (solid) and $N=4$ (dashed) emitters.
        (a) Intracavity photon number starting from a well defined value $N/2$ with $\kappa/g = 0.1$, $\gamma/g = 0.1$.
        (b) Structure factor witness $\ew{W}$ clearly  detecting entangled states close to the half-inverted Dicke state as indicated by the expectation value $\ew{D_N} = \braket{ D_{N, N/2} | \rho_\mathrm{q} | D_{N, N/2}}$.
        (c) Minimum value for $\ew{W}$ as a function of emitter decay rate $\gamma/g$ after initializing the system in the photon state as in (a).
    }
    \label{fig:intial_photon_fock_state}
\end{figure}

\paragraph*{Coherent Dicke-state preparation.}---Advances in quantum technology, and in particular cavity QED, make it possible to create well-defined photon states \cite{bertet_generating_2002,hofheinz_generation_2008, cooper_experimental_2013}.
As we show, this opens the door for experiments that use Fock-state quantum light to generate entangled Dicke states in the dynamics of an emitter-cavity system.
In Fig.~\ref{fig:intial_photon_fock_state} we show the dynamics of $N=2$ (solid lines) and $N=4$ (dashed) emitters strongly interacting with the cavity mode.
As initial conditions, we assume Fock states of $n_p=\frac{N}{2}$ photons in the cavity.
During the time evolution shown in Fig.~\ref{fig:intial_photon_fock_state}(a), one clearly sees the detection of entanglement by the structure factor witness $\ew{W}$ (black lines Fig.~\ref{fig:intial_photon_fock_state}(b)).
Panel (b) further reveals the intermittent creation of the corresponding half-inverted Dicke state $\ket{D_{N, N/2}}$ (orange lines).
The deterministic creation of the state $\ket{D_{N,N/2}}$ by the coherent dynamics can be explained in a simplified model that represents a generalization of the Jaynes-Cummings model to a multi-level emitter coupled to a cavity, where the emitter levels are represented by the Dicke states $\ket{N/2,m}$.
This is due to the fact that the Hamiltonian \eqref{eq:H_RWA} commutes with the operator $\bm{J}^2$ and, thus, conserves the quantum number $j$.

The creation of Dicke-state entanglement in a deterministic fashion hinges on the ability to control the photon state in the cavity, and on the suppression of decoherence effects.
To show this, we explicitly consider emitter dephasing over the range from strong to weak light-matter coupling and show the attainable entanglement in terms of the minimal value of $\ew{W}$ in Fig.~\ref{fig:intial_photon_fock_state}(c).
In the strong-coupling regime, the emitters are truly found to be in an entangled state at the maximum of their excitation by the cavity field.
With increasing dissipation rate $\gamma$, the overlap with the targeted Dicke state and the resulting entanglement reduces.
At a single-emitter dissipation rate larger than $\gamma/g = 1.32$ (0.91) for $N=2$ ($N=4$) emitters, entanglement is no longer detected.
As expected, the genuine multipartite entanglement becomes less robust to decoherence effects as the number of emitters increased.

\paragraph*{Conclusion.}
---Dicke super- and subradiance is a recurrent topic in fundamental, atom- and semiconductor physics.
In recent observations, e.g.~in semiconductor nanolasers, superradiant emission bursts have been observed.
With a maximum emission-rate enhancement at half inversion and the well-known scaling behavior for the half-inverted Dicke state $\propto N^2$, there is no reason not to assume that this type of emission-time enhancement arises from a transitory Dicke state.
We show that, contrary to this expectation, signatures of superradiance are not linked to the existence of Dicke states, but result from a mixture of (pseudo-) angular momentum eigenstates with both sub- and superradiant contributions.

By combining methods from semiconductor optics and quantum information theory, we could establish a close connection between emitter correlations that modify spontaneous emission, and a witness for detecting Dicke-state entanglement.
While the emission rate enhancement benefits from a combinatorial scaling factor of all emitter pairs in an ensemble, entanglement does not, which is why entanglement is a much harder criterion to fulfill, and is ultimately not found even in the presence of strong emission-time enhancement.

Finally, we propose to use Fock-state quantum light to excite a cavity QED system, which can generate a transitory entangled Dicke state that can be discovered by an experimentally accessible entanglement witness.
Then, the emission of superradiant light truly becomes a signature of multipartite entanglement within the emitter ensemble.
We believe that this proposal has broad applicability and could lead to exciting new avenues of research in the field.

\begin{acknowledgments}
    C.~Gies is grateful for funding from the DAAD (Deutscher Akademischer Austauschdienst), facilitating a research sabbatical at the University of Otago and the Dodd Walls Centre.
F.~Lohof  acknowledges funding from the central research development fund (CRDF) of the University of Bremen.
\end{acknowledgments}

\bibliography{biblio_SR_letter.bib}

\end{document}